\documentclass[pra,twocolumn,showpacs]{revtex4}
\usepackage{amsmath,amssymb,mathrsfs}
\usepackage{hyperref}
\usepackage{psfrag}
\usepackage{graphicx}
\usepackage{bm}
\usepackage{verbatim,color,ulem}

\newcommand{\beq}{\begin{equation}}
\newcommand{\eeq}{\end{equation}}
\newcommand{\beqa}{\begin{eqnarray}}
\newcommand{\eeqa}{\end{eqnarray}}

\renewcommand{\vec}{\mathbf}

\begin{document}

\title{Tunneling dynamics of bosonic Josephson junctions assisted by a cavity field}
\author{G. Szirmai$^{1}$, G. Mazzarella$^{2}$, L. Salasnich$^{2}$}
\affiliation{$^{1}$Institute for Solid State Physics and Optics -
Wigner Research Centre for Physics,
Hungarian Academy of Sciences, H-1525 Budapest P.O. Box 49, Hungary
\\
$^{2}$Dipartimento di Fisica e Astronomia ``Galileo Galilei''
and CNISM, Universit\`a di Padova, Via Marzolo 8, I-35131 Padova, Italy}

\date{\today}

\begin{abstract}
We study the interplay between the dynamics of a Bose-Einstein condensate in a double-well potential and that of an optical cavity mode. The cavity field is superimposed to the double-well potential and affects the atomic tunneling processes. The cavity field is driven by a laser red detuned from the bare cavity resonance; the dynamically changing spatial distribution of the atoms can shift the cavity in and out of resonance. At resonance the photon number is hugely enhanced and the atomic tunneling becomes amplified. The Josephson junction equations are revisited and the phase diagram is calculated. We find new solutions with finite imbalance and at the  same time a lack of self-trapping solutions due to the emergence of a new separatrix resulting from enhanced tunneling.
\end{abstract}


\maketitle

\section{Introduction} 

Bosonic realization of the Josephson effect with ultracold atoms in symmetric double-well potentials \cite{raghavan1999coherent,andrews1997observation,wang2005atom,albiez2005direct,esteve2008squeezing,leblanc2011dynamics,mb,gio1,gio2,gio3,gio4,gio5,gio6,julia2010bose,julia2010macroscopic} is a paradigm of a few degree of freedom system which is described by the nonlinear equations of motion characteristic to a nonrigid pendulum. The coherent dynamics of the matter wave exhibits Josephson oscillations or self-trapping depending on the tunneling and interaction parameters of the gas sample.
Recently we have shown that an additional laser pump tightly focused to the center of the double-well barrier can be used to tune, or eventually, change the sign of the tunneling amplitude between the left and right valleys \cite{szirmai2014effect}.

When an ensemble of ultracold atoms is placed inside a high-Q optical resonator, atom-light interaction plays a crucial role, even in the dilute gas limit, due to the multiple recycling of the cavity photons from the mirrors \cite{ritch2013cold}. As an immediate consequence, there is a significant mutual back-action between the atomic and photon degrees of freedom. The optical potential created by the cavity photons alters significantly the external trapping potential, which modifies the atomic motion and thus the dispersive medium where the photon field propagates.  The parameters of the low-energy description of the atomic motion become dependent on the state of the photon filed being itself a dynamic quantitiy \cite{maschler2005cold,larson2008mott,larson2008quantum,maschler2008ultracold,vukics2009cavity,niedenzu2010microscopic,fernandez2010quantum,li2013lattice}. Addition of an optical resonator to the double-well setup therefore leads to a hybrid system where some or all of the bosonic Josephson junction parameters become dynamical quantities. The purpose of the present paper is to study the modification of the coherent dynamics of a bosonic Josephson junction in the presence of the dynamical optical potential created by the cavity. Earlier proposals \cite{corney1998homodyne,vukics2007microscopic,zhang2008cavity,zhang2008mean,zuppardo2014cavity} already considered the effect of the cavity field on the on-site energies of the junction. A recent proposal by Larson and Martikainen considered the extreme case where the double-well barrier was formed by the cavity field \cite{larson2010ultracold}. Here, due to a modified geometry, we also consider the phenomena arising from cavity assisted boson tunneling.

Similar systems, where the two-mode double-well setup is extended with an additional bosonic mode include bosonic Josephson junctions in the presence of an impurity \cite{rinck2011effects,mulansky2011impurity,mumford2014impurity}, or the Dicke model realized with bosonic superfluids inside optical cavities \cite{emary2003chaos,baumann2010dicke,nagy2010dicke}. It is common in these systems that the atomic dynamics represents a bosonic Josephson junction, which can be described semiclassicaly with a non-rigid pendulum, while the impurity or cavity field is described by an additional oscillator. Recently, the behavior of superfluid fermions in double well potentials \cite{zhou2014josephson} and the superradiant transition of fermions in the field of a single mode cavity are also investigated \cite{keeling2014fermionic,piazza2014umklapp,chen2014superradiance}. These systems realize thus the first step starting from a simple scenario of a quantum  oscillator towards the complications in many-body quantum physics.

\section{Model}

\begin{figure}[tb!]
\centering
\includegraphics{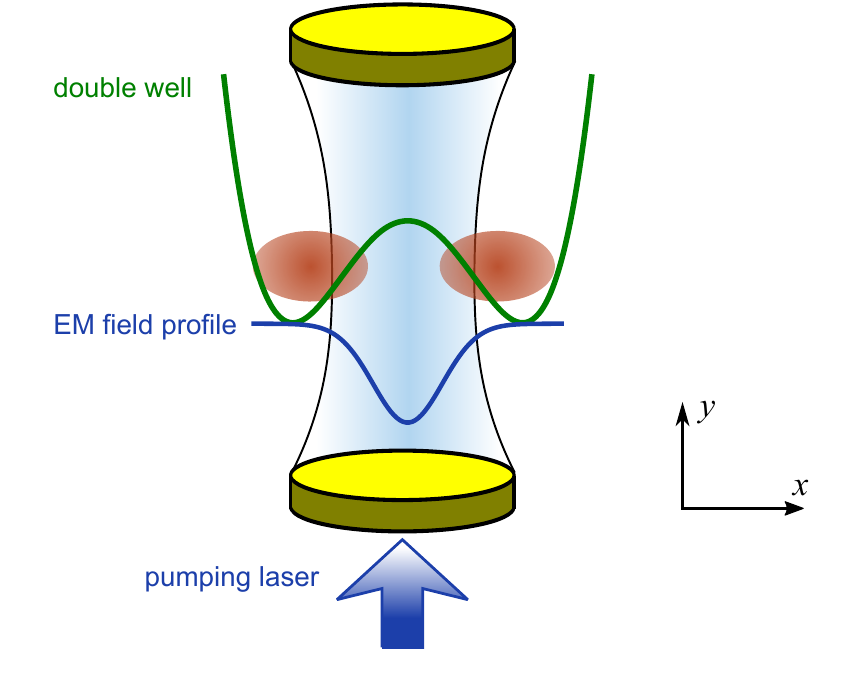}
\caption{(Color online) The illustration of the setup. 
The bosonic Josephson Junction is created by magnetic or optical means 
along the $x$ direction. A Fabry-P\'erot cavity is placed around the 
junction with an axis orthogonal to the junction. The resonator is 
operated on the TEM$_{00}$ mode. }
\label{fig:scheme}
\end{figure}

We consider a bosonic Josephson junction (BJJ) formed by an interacting Bose-Einstein condensate in a symmetric double-well potential \cite{raghavan1999coherent}. The BJJ is placed inside a single mode, high-Q Fabry-P\'erot resonator, with the cavity axis perpendicular to the direction of the double-well potential. We assume that the cavity is pumped by a laser with frequency $\omega_L$ and amplitude $\eta$ through one of its mirrors and is operated at the $\mathrm{TEM}_{00}$ mode. The peak value of the cavity field is at the center of the double-well barrier. The setup is illustrated in Fig. \ref{fig:scheme}. The Bose-Einstein condensate is formed by bosonic alkaline or alkaline-earth-metal atoms having an electronic  ground state and an optically accessible excited state with a transition frequency $\omega_A$.   The atoms are confined by a double-well (DW) potential,  $V_{DW}(x)$, along the $x$ direction. We assume that the confinement is tightly harmonic with frequency $\omega_H$ in the $(y,z)$ plane, i.e. the total external potential $V({\bf r})$ for the atoms is given by 
\beq 
\label{eq:pot}
V(\vec{r}) = V_{DW}(x) + {1\over 2} m \omega_{H}^2 (y^2+z^2) \; ,
\eeq 
where $m$ is the mass of the bosons. In this way, the atomic Hamiltonian takes the form
\begin{multline}
\label{eq:Hamatom}
{\hat H}_A=\sum_{\sigma=g,e}\int d^3{\bf r}\ 
{\hat \Psi}_\sigma^\dagger(\vec{r})\bigg[-\frac{\hbar^2}
{2m}\nabla^2+V(\vec{r})
+\hbar\omega_{A}\ \delta_{\sigma,e}\\
+\frac{1}{2}\sum_{\sigma'}g_{\sigma\sigma'}{\hat \Psi}_{\sigma'}^\dagger(\vec{r})
{\hat \Psi}_{\sigma'}(\vec{r})\bigg]{\hat \Psi}_{\sigma}(\vec{r}).
\end{multline}
The atomic transition frequency is in the optical range, i.e. $\omega_A\approx10^{15}\,\mathrm{Hz}$. The trapping 
potential $V(\vec{r})$ is created either by purely optical means or a 
combination of magnetic trapping and laser blockade. In principle it can 
depend on the internal state of the atom, but it will play no role in our 
considerations. For example, with alkaline-earth-metal atoms a state 
independent potential can be realized by using lasers at the magic wavelength, 
where the polarizability of the $g$ and $e$ states are equal 
\cite{gerbier2010gauge}. Atom-atom interaction is modelled by s-wave 
scattering, with coupling constants $g_{gg}$, $g_{ee}$ and $g_{ge}$, 
arranged into $g_{\sigma,\sigma'}=4\pi\hbar^2a_{\sigma,\sigma'}/m$. Here $a_{\sigma,\sigma'}$ 
is the s-wave scattering length pertaining to collisions between two atoms in the state $\sigma$ and $\sigma'$, respectively.

The above bosonic Josephson Junction is assumed to be inside a high-Q 
optical cavity with a characteristic frequency $\omega_C$ close to the 
atomic transition frequency $\omega_A$. The cavity is pumped through one 
of the mirrors by a coherent laser light. The single mode of the radiation 
inside the cavity is described by the Hamiltonian \cite{ritch2013cold}
\begin{equation}
\label{eq:Hamcav}
{\hat H}_C=\hbar\omega_C\, {\hat a}^\dagger \, 
{\hat a} -i\hbar\eta\bigg(e^{i\omega_L t} {\hat a} - 
e^{-i\omega_L t} {\hat a}^\dagger\bigg),
\end{equation}
where $\eta>0$ is the strength 
of the driving laser, and $\omega_L$ is its single-mode frequency. 
The cavity axis is chosen along the $y$ 
direction, and the cavity TEM$_{00}$ mode function has the form 
\beq 
f(\vec{r})= \sqrt{2\over L} \,\cos(k\,y) \, 
{e^{-(x^2+z^2)/(2\sigma^2)}\over \pi^{1/2} \sigma},
\eeq 
with $k=\omega_C/c$ is the wave number of the cavity mode, 
$L$ is the distance between the mirrors 
and $\sigma$ is the width of the Gaussian profile in the $(x,z)$ plane.

Atom-light interaction is described by the standard Jaynes-Cummings interaction
\begin{equation}
{\hat H}_I=-i\hbar \Omega_R\int d^3{\bf r}\ f(\vec{r}) \big[{\hat a}\,
{\hat \Psi}_e^\dagger(\vec{r})
{\hat \Psi}_g(\vec{r})-{\hat a}^\dagger\,
{\hat \Psi}_g^\dagger(\vec{r}){\hat \Psi}_e(\vec{r})\big],
\end{equation}
which is valid in the rotating wave approximation 
\cite{standardquantumopticsbook}. The parameter $\Omega_R$ is the 
single-photon Rabi frequency. The full Hamiltonian is assumed to be the sum of the individual 
contributions ${\hat H}={\hat H}_A+{\hat H}_C+{\hat H}_I$. In order to eliminate the time dependence 
from $H_C$ we switch to a frame rotating with $\omega_L$ with the help 
of the following unitary transformation,
\begin{equation}
\label{eq:rotatingframe}
{\hat U}(t)=\exp\bigg\lbrace i\omega_L t \bigg[{\hat a}^\dagger {\hat a} + 
\int d^3{\bf r} \ 
{\hat \Psi}_e^{\dagger}(\vec{r}){\hat \Psi}_e(\vec{r})\bigg]\bigg\rbrace.
\end{equation}
In the rotating frame the annihilation operators are replaced by 
${\hat a}\rightarrow {\hat U}\,{\hat a}\,{\hat U}^\dagger$, 
${\hat \Psi}_e(\vec{r})\rightarrow {\hat U}\,
{\hat \Psi}_e(\vec{r})\,{\hat U}^\dagger$, and the 
Hamiltonian is transformed to 
${\hat H}\rightarrow {\hat U}\,{\hat H}\,{\hat U}^\dagger + 
i\hbar(\partial_t {\hat U}){\hat U}^\dagger$.
The Hamiltonian in the rotating frame now reads as
\begin{multline}
\label{eq:fullham}
{\hat H}=\sum_{\sigma=g,e}\int d^3{\bf r}\ {\hat \Psi}_\sigma^\dagger(\vec{r})
\bigg[-\frac{\hbar^2}{2m}\nabla^2+V(\vec{r})
-\hbar\Delta_A\delta_{\sigma,e}\\
+\frac{1}{2}\sum_{\sigma'}g_{\sigma\sigma'}{\hat \Psi}_{\sigma'}^\dagger(\vec{r})
{\hat \Psi}_{\sigma'}(\vec{r})\bigg]{\hat \Psi}_{\sigma}(\vec{r})
-\hbar\Delta_C\, {\hat a}^\dagger\,{\hat a}-i\hbar\eta\big({\hat a} 
- {\hat a}^\dagger\big)\\
-i\hbar \Omega_R\int d^3{\bf r}\ f(\vec{r}) \big[{\hat a}\,
{\hat \Psi}_e^\dagger(\vec{r})
{\hat \Psi}_g(\vec{r})-{\hat a}^\dagger\,{\hat \Psi}_g^\dagger(\vec{r})
{\hat \Psi}_e(\vec{r})\big].
\end{multline}
The time dependence has disappeared, but we have the following detunings 
instead of the bare frequencies, $\Delta_A=\omega_L-\omega_A$, 
$\Delta_C=\omega_L-\omega_C$. The Heisenberg equations of the field 
operators are given by
\begin{subequations}
\begin{multline}
i\hbar\partial_t {\hat a}=[{\hat a},{\hat H}]\\
=-\hbar \Delta_C \, {\hat a} 
+i\hbar \eta
+ i\hbar \Omega_R\int d^3{\bf r} f(\vec{r}){\hat \Psi}_g^\dagger(\vec{r})
{\hat \Psi}_e(\vec{r}),
\end{multline}
\begin{multline}
i\hbar\partial_t {\hat \Psi}_g(\vec{r})=[{\hat \Psi}_g(\vec{r}),{\hat H}]
=\Bigg[-\frac{\hbar^2}{2m}\nabla^2+V(\vec{r})\\
+\sum_\sigma g_{g\sigma}{\hat \Psi}_\sigma^\dagger(\vec{r})
{\hat \Psi}_\sigma(\vec{r})\Bigg]
{\hat \Psi}_g(\vec{r})+i\hbar \Omega_R f(\vec{r}){\hat a}^\dagger\,
{\hat \Psi}_e(\vec{r}),
\end{multline}
\begin{multline}
\label{eq:eqmoexc}
i\hbar\partial_t {\hat \Psi}_e(\vec{r})=[{\hat \Psi}_e(\vec{r}),{\hat H}]=
\Bigg[-\frac{\hbar^2}{2m}\nabla^2+V(\vec{r})-\hbar\Delta_A\\
+\sum_\sigma g_{e\sigma}{\hat \Psi}_\sigma^\dagger(\vec{r})
{\hat \Psi}_\sigma(\vec{r})\Bigg]{\hat \Psi}_e(\vec{r})-
i\hbar \Omega_R f(\vec{r}) {\hat a} \,
{\hat \Psi}_g(\vec{r}).
\end{multline}
\end{subequations}

In the limit, when the atomic detuning is much larger than the other 
characteristic frequency scales of the system, $\Delta_A\gg\sqrt{N}\Omega_R,
\hbar^{-1}V, \hbar^{-1}N g_{\sigma,\sigma'}$, the population of the excited 
state is small and follows adiabatically the ground state. The excited state 
field operator thus can be eliminated. Under these conditions the square 
bracket in Eq. \eqref{eq:eqmoexc} is dominated by the detuning, and 
the excited state field operator immediately relaxes to its equilibrium value,
\begin{equation}
{\hat \Psi}_e(\vec{r})\simeq -i\frac{\Omega_R f(\vec{r})}{\Delta_A}a\,
{\hat \Psi}_g(\vec{r}).
\end{equation}
The resulting dynamics of the photon field and the ground state atoms 
are given by
\begin{subequations}
\label{eqs:eqmoeliminated}
\begin{equation}
i\hbar\partial_t {\hat a}=-\hbar \Delta_C \,{\hat a}+i\hbar \eta
+ \hbar \frac{\Omega_R^2}{\Delta_A} {\hat a}\int d^3 r f^2(\vec{r})
{\hat \Psi}_g^\dagger(\vec{r}){\hat \Psi}_g(\vec{r}),
\end{equation}
\begin{multline}
i\hbar\partial_t {\hat \Psi}_g(\vec{r})=\Bigg[-\frac{\hbar^2}{2m}\nabla^2+
V(\vec{r})+\hbar\frac{\Omega_R^2}{\Delta_A}f^2(\vec{r})
{\hat a}^\dagger {\hat a}\\
+g_{gg}{\hat \Psi}_g^\dagger(\vec{r}){\hat \Psi}_g(\vec{r})\Bigg]
{\hat \Psi}_g(\vec{r}) ,
\end{multline}
\end{subequations}
where we have neglected interaction with excited state atoms, since the 
population of the excited state is much smaller than that of the ground state. 
Equations \eqref{eqs:eqmoeliminated} can be obtained from the following 
effective Hamiltonian
\begin{multline}
\label{eq:Hamelim}
{\hat H}_{\text{eff}}=-\hbar\Delta_C {\hat a}^\dagger {\hat a} - 
i\hbar\eta( {\hat a}-{\hat a}^\dagger)+
\int d^3{\bf r}\ {\hat \Psi}_g^\dagger(\vec{r})\Bigg[-\frac{\hbar^2}{2m}
\nabla^2\\
+V(\vec{r})+\hbar U_0\,{\hat a}^\dagger {\hat a} 
f^2(\vec{r})+ \frac{1}{2} g_{gg} 
{\hat \Psi}_g^\dagger(\vec{r}) {\hat \Psi}_g(\vec{r})\Bigg] 
{\hat \Psi}_g(\vec{r}),
\end{multline}
where we have introduced $U_0=\Omega_R^2/\Delta_A$. 
As a result of the adiabatic  elimination of the excited state, atom-photon interaction now is the dispersive photon scattering on the atoms. As a consequence a new optical potential has appeared, 
with position dependence $f^2(\vec{r})$ and an effective amplitude 
$U_0\,a^\dagger a$. Therefore when the atomic transition is red detuned 
from the pumping ($U_0<0$), the atoms are attracted to the intensity 
maxima of the cavity field, which effectively lowers the double-well 
barrier. This effect is proportional to the photon number, thus the state 
of the cavity dynamically influences the parameters of the Josephson 
Junction.

In order to arrive to a simplified, two-mode description of the system we assume 
that the unperturbed double-well potential Eq. \eqref{eq:pot} defines local Wannier-like states $w_1(x)$ and $w_2(x)$ centered around the minima of the unperturbed double-well potential $V_{DW}(x)$, which stay practically unchanged when turning the cavity field on. In practice, 
this assumption means that the third lowest energy eigenstate in the double-well potential stays away from the low-energy doublet (the energy of the 
symmetric and antisymmetric combinations of Wannier orbits) even when a 
classical cavity field is present. By denoting the energy difference 
between the third lowest energy state and  the lowest energy doublet by 
$\Delta_{DW}$, we have the condition $\Delta_{DW}\gg -U_0 
\xi^2\langle f^2(\vec{r})\rangle$, where $\xi^2$ is the average number of photons in the cavity and the average $\langle f^2(\vec{r})\rangle$ is calculated with respect to the condensate wave function. 
The atomic field operator is approximated as
\begin{equation}
\label{eq:wannierexp}
{\hat \Psi}_g(\vec{r})= 
\left( w_1(x)\ {\hat b}_1+w_2(x)\ {\hat b}_2 \right) 
{e^{-(y^2+z^2)/(2l_H^2)} \over \pi^{1/2} l_H} \; ,
\end{equation}
where $\hat b_1$ and $\hat b_2$ are the bosonic annihilation operators of 
the Wannier-like functions centered around the two 
minima of the DW potential and $l_H=\sqrt{\hbar/(m\omega_H)}$ is the 
characteristic length of the strong harmonic confinement 
in the $(y,z)$ plane. 

\begin{figure}[t!]
\centering
\includegraphics[scale=0.65]{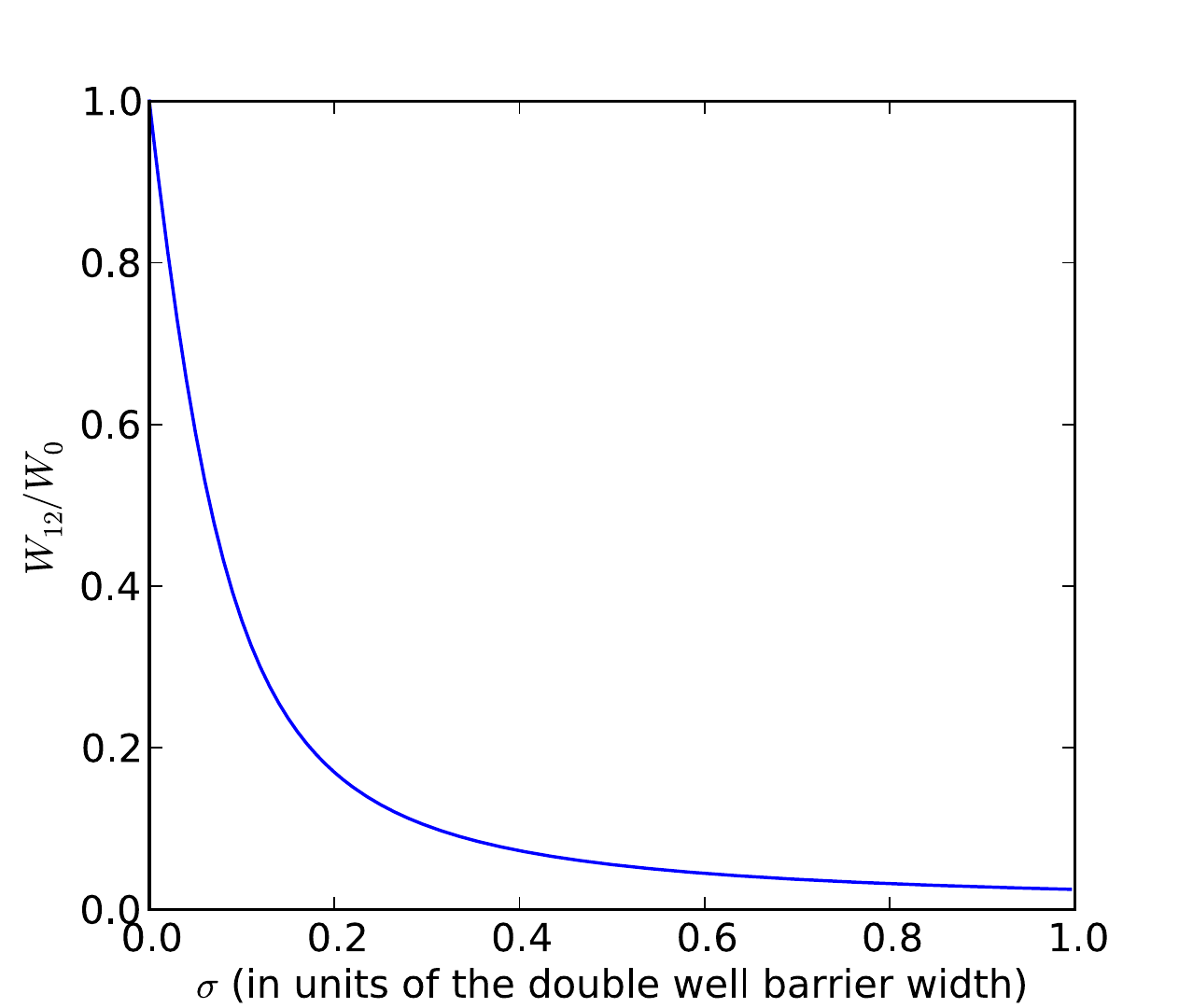}
\caption{The ratio between $W_{12}$ and $W_0$ as a function of the ratio between the width of the cavity waist and the width of the double-well barrier. The parameters are chosen in such a way that the Wannier functions have some small but non-zero overlap ($J\approx0.1\epsilon$).}
\label{fig:ratio}
\end{figure}

By substituting Eq. \eqref{eq:wannierexp} into the effective Hamiltonian Eq. \eqref{eq:Hamelim}, and since the double well is symmetric, one arrives to the Hamiltonian
\beq
{\hat H} = {\hat  H}_L +{\hat  H}_J +{\hat  H}_{JL},
\label{ham}
\eeq
where the bare cavity is described by
\beq
{\hat H}_L = - \hbar\Delta_C  \,\hat a^\dagger\hat a  - i\hbar\eta
\left( {\hat a} - {\hat a}^{\dagger} \right) .
\eeq
The bosonic Josephson junction is described by
\beq
{\hat H}_J = \epsilon \, {\hat  N}_A - J \left( {\hat b}_1^{\dagger} 
{\hat b}_2 + {\hat b}_2^{\dagger} {\hat b}_1 \right)
+ {U\over 2} \left( \hat b_1^\dagger\hat b_1^\dagger\hat b_1\hat b_1+\hat b_2^\dagger\hat b_2^\dagger\hat b_2\hat b_2\right) \; ,
\eeq
where ${\hat N}_A = {\hat b}_1^{\dagger}{\hat b}_1 + {\hat b}_2^{\dagger}{\hat b}_2$ is the total number of atoms. The parameter $\epsilon$ is the on-site energy of a single well, $J$ represents the tunneling amplitude, and $U$ is the on-site interaction energy. The dispersive interaction between the atoms and the cavity photons is modeled by
\beq
{\hat H}_{JL} = {\hat N}_L \left[W_0 \hat N_A + W_{12} ({\hat b}_1^\dagger {\hat b}_2 + {\hat b}_2^\dagger {\hat b}_1) \right],
\eeq
with ${\hat N}_L=\hat a^\dagger\hat a$ the photon number. The parameters $W_0$, $W_{12}$ are the AC-Stark shift and the cavity assisted tunneling amplitude, respectively \cite{standardquantumopticsbook}. For red detuned atoms both $W_0$ and $W_{12}$ are negative. Thus the cavity field shifts the on-site energies downwards and assists the tunneling by increasing the effective tunneling rate. The magnitude of the parameter $W_0$ is always bigger than that of $W_{12}$. The advantage of this setup is that it can be achieved that $W_0$ and $W_{12}$ are close to each other [see Eqs. \eqref{eq:W0}, \eqref{eq:W12}  and Fig. \ref{fig:ratio}].

The parameters are the single well energy,
\begin{subequations}
\label{eqs:hubparams}
\begin{equation}
\label{eq:eps}
\epsilon=\hbar \omega_H+\int dx \,w_j^*(x)\bigg[-\frac{\hbar^2}{2m}{d^2\over dx^2}
+V_{DW}(x)\bigg]w_j(x) ,
\end{equation}
which is independent of $j$ for a symmetric DW potential. 
The hopping (tunneling) energy, 
\begin{equation}
\label{eq:J}
J=-\int dx\,w_1^*(x)\bigg[-\frac{\hbar^2}{2m}{d^2\over dx^2} + 
V_{DW}(x) \bigg]w_2(x) 
\end{equation}
can be chosen to be real and positive. Finally, the on-site interaction 
strength is given by
\begin{equation}
\label{eq:U}
U={g_{gg}\over 2\pi l_H^2} \int dx \, |w_j(x)|^4.
\end{equation}

Parameters of the atom-light interaction part of the Hamiltonian are the following. First, the AC-Stark shift is given by
\beq
\label{eq:W0}
W_0 = \frac{\hbar U_0\Big(1+e^{-k^2 l_H^2}\Big)}{L \pi \sigma \sqrt{l_H^2+\sigma^2}} \int dx  |w_j(x)|^2 \, e^{-x^2/\sigma^2} , 
\eeq
while the cavity assisted tunneling constant is
\beq
\label{eq:W12}
W_{12}=\frac{\hbar U_0 \Big(1+e^{-k^2 l_H^2}\Big)}{L \pi \sigma \sqrt{l_H^2+\sigma^2}} \int dx  w_1^*(x)\, w_2(x) \, e^{-x^2/\sigma^2} .  
\eeq
\end{subequations}
In the most important case, when the pumping is red detuned from the atomic transition ($U_0<0$), the parameters $W_0,W_{12}<0$. In other words the cavity field shifts the on-site energies downwards and assists the tunneling by increasing the effective tunneling rate.
The overlap between different Wannier functions is never perfect (in fact, it is usually very small), therefore $|W_{12}|<|W_0|$. In previous proposals, for different setups \cite{zhang2008cavity,zhang2008mean}, the value of $W_{12}$ was negligible. Here, due to the transverse direction of the cavity, the ratio between $W_{12}$ and $W_0$ depends on the width $\sigma$ of the $\mathrm{TEM}_{00}$ mode function. When the cavity waist is much smaller than the width of the double-well barrier the Gaussian factor inside the integrals Eqs. \eqref{eq:W0} and \eqref{eq:W12} basically samples only the part where the Wannier functions overlap, and therefore $W_{12}$ can even be close $W_0$. It is illustrated in Fig. \ref{fig:ratio}.

It is straightforward to derive the Heisenberg equations of motion for the operators $\hat{b}_j$ ($j=1,2$) and $\hat{a}$. First we drop the on-site energy term $\epsilon \hat N_A$ from the Hamiltonian, since $\hat N_A$ is a constant of motion. Then, 
\begin{subequations}
\label{heisenberg}
\begin{align}
i\hbar \, {d\over dt} \hat{b}_1 &= W_0\,\hat{a}^{\dagger}\hat{a}\,\hat{b}_1-(J-W_{12}\,\hat{a}^{\dagger}\hat{a})\hat{b}_2
+ U\,\hat{n}_1\hat{b}_1,\\
i\hbar \, {d\over dt} \hat{b}_2 &=  W_0\,\hat{a}^{\dagger}\hat{a}\,\hat{b}_2-(J-W_{12}\,\hat{a}^{\dagger}\hat{a})\hat{b}_1
+ U\,\hat{n}_2\hat{b}_2,\\
i\hbar \, {d\over dt} \hat{a} &= -[\hbar\Delta_C -W_0\hat N_A-W_{12}(\hat{b}^{\dagger}_1
\hat{b}_2+\hat{b}^{\dagger}_2\hat{b}_1)]\hat{a}+i\hbar\eta,
\end{align}
\end{subequations}
with $\hat n_j=\hat b_j^\dagger \hat b_j$, the atomic population in the well $j$.

\textit{Mean-field approximation -} Under the semiclassical (mean-field) approximation the system is
in the full coherent state (FCS)
\beq
|FCS\rangle = |\beta_1 \rangle_A \otimes |\beta_2\rangle_A
\otimes |\alpha\rangle_L
\eeq
such that ${\hat b}_j|\beta_j\rangle_A = \beta_j |\beta_j\rangle_A$,
where $\beta_j=\sqrt{N_j(t)}\, e^{i\theta_j(t)}$ with $N_j(t)$ the
average number of atoms at the $j$th well at time $t$ and
$\theta_j(t)$ the corresponding phase. Similarly
$\hat{a}|\alpha\rangle_L =\alpha |\alpha\rangle_L$, where
$\alpha= \xi(t) \, e^{i\phi(t)}$ with $N(t)=\xi(t)^2$ the average
number of photons in the cavity at time $t$ and $\phi(t)$
the corresponding phase.
We obtain therefore the following ordinary differential equations (ODEs):
\begin{subequations}
\label{odes}
\beqa
\dot{z} &=&-2\nu\sqrt{1-z^2}\sin\theta,\label{eq:odez}\\
\dot{\theta} &=& \left(\tilde g + \frac{2\,\nu}{\sqrt{1-z^2}}\cos{\theta}\right)z,\label{eq:odetheta}
\\
\dot{\xi} &=& \eta\,\cos{\phi},\label{eq:odexi}
\\
\dot{\phi} &=& \delta_C - \frac{\eta}{\xi}\sin{\phi},\label{eq:odephi}
\eeqa
\end{subequations}
where $N_A=N_1(t)+N_2(t)$ is the total atom number, $z(t)=(N_1(t)-N_2(t))/N_A$ is
the fractional imbalance of the atomic population, and $\theta(t)=\theta_2(t)-\theta_1(t)$ is
the atomic relative phase. We have also introduced here the following parameters with frequency dimensions: $\tilde g=UN_A/\hbar$, which is the mean-field frequency shift due to atomic collisions; $\nu=(J-W_{12}\xi^2)/\hbar$, standing for the effective tunneling strength modified by the photon assisted process; $\delta_C=\Delta_C-N_A(W_0+W_{12}\sqrt{1-z^2}\cos\theta)/\hbar$ is the effective cavity detuning. Notice that in the absence of radiation fields, i.e. $\xi(t)=\phi(t)=0$ at any time $t$, the above ODEs reduce to those of the standard BJJ dynamics \cite{raghavan1999coherent}.  Also notice, that the parameters $\nu(t)$ and $\delta_C(t)$ are shorthand notations and depend on the mean-field variables. Moreover, the role of the hopping $J$ in the bare BJJ pertains now to the time-dependent term $\nu(t)$.

\section{Fixed points}

\begin{figure}[tb!]
\centering
\includegraphics[scale=0.6]{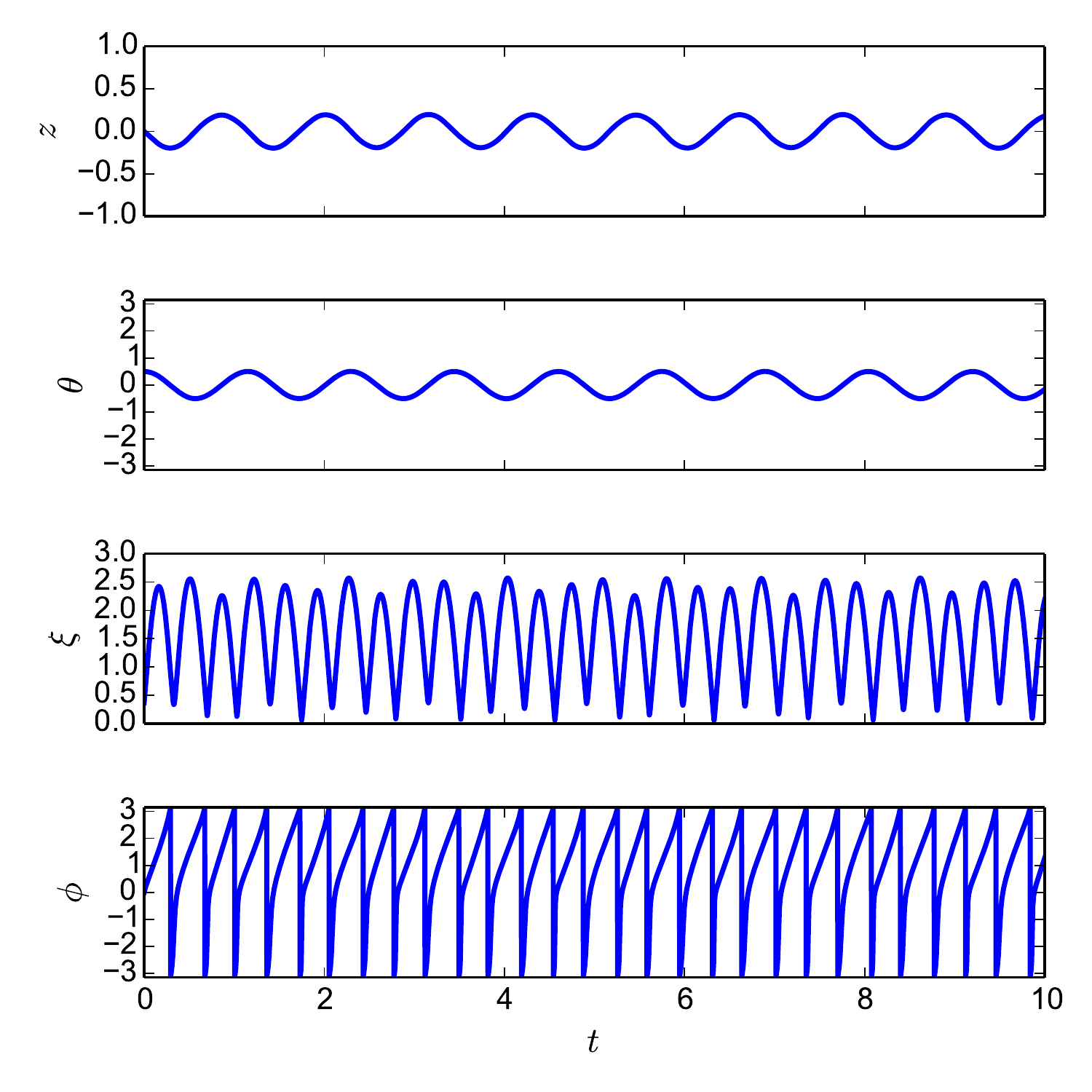}
\caption{(Color online) Time evolution of the variables $(z,\theta,\xi,\phi)$ around the fixed point ${\bf X}_2$. Time is measured in units of $\hbar/J$. The time evolution exhibits Josephson oscillations of the system analogous to that of pure BJJs.  The four panels correspond to the four components of the state vector ${\bf X}$. The parameters are: $\hbar\Delta_C=-100J$, $W_0 N_A=-90 J$, $W_{12} N_A=-30 J$, $U N_A=12 J $, $N_A=1000$, $\hbar\eta=20 J$. The initial condition is ${\bf X}(t=0)=(0,0.5,0,0)$.}
\label{fig:Josc}
\end{figure}

Let us use a more concise vector notation, ${\bf X}= (z,\theta,\xi,\phi)$, and discuss the fixed points of the autonomous ODEs (\ref{odes}), with $d{\bf X}/dt=0$. First consider the \textit{zero imbalance equilibrium} solutions with $z=0$. 
In this case $\dot{\theta}=0$. The condition $\dot{z}=0$ can be
met when $\theta=0$, or $\theta=\pi$. Requiring $\dot{\xi}=0$ 
leads to $\phi=\pm\pi/2$. Finally, the condition $\dot{\phi}=0$ 
allows one to find $\xi$ and write down the first four stationary points:
\begin{subequations}
\begin{align}
\label{fpzero1}
{\bf X}_{1}&=\left(0,0,\frac{\hbar\eta}{\hbar\Delta_C-N_A(W_0+W_{12})},\frac{\pi}{2}\right),\\
{\bf X}_{2}&=\left(0,0,\frac{\hbar\eta}{N_A(W_0+W_{12})-\hbar\Delta_C},-\frac{\pi}{2}\right),\\
{\bf X}_{3}&=\left(0,\pi,\frac{\hbar\eta}{\hbar\Delta_C-N_A(W_0-W_{12})},\frac{\pi}{2}\right),\\
{\bf X}_{4}&=\left(0,\pi,\frac{\hbar\eta}{N_A(W_0-W_{12})-\hbar\Delta_C},-\frac{\pi}{2}\right).
\end{align}
\end{subequations}
Notice that $\xi$ is the amplitude of the photon field, consequently, $\xi>0$. Therefore, for $\delta_C>0$ we have ${\bf X}_{1}$ and ${\bf X}_{3}$; while for $\delta_C<0$ we have ${\bf X}_{2}$ and ${\bf X}_{4}$ fixed points.
Oscillations around these fixed points correspond to plasma oscillations of the bare double-well setup as illustrated in Fig. \ref{fig:Josc}. Here the system starts from the following initial condition: $z=0$, $\theta=0.5$, $\xi=0$, $\phi=0$.
\begin{figure}[!tb]
\centering
\includegraphics[scale=0.6]{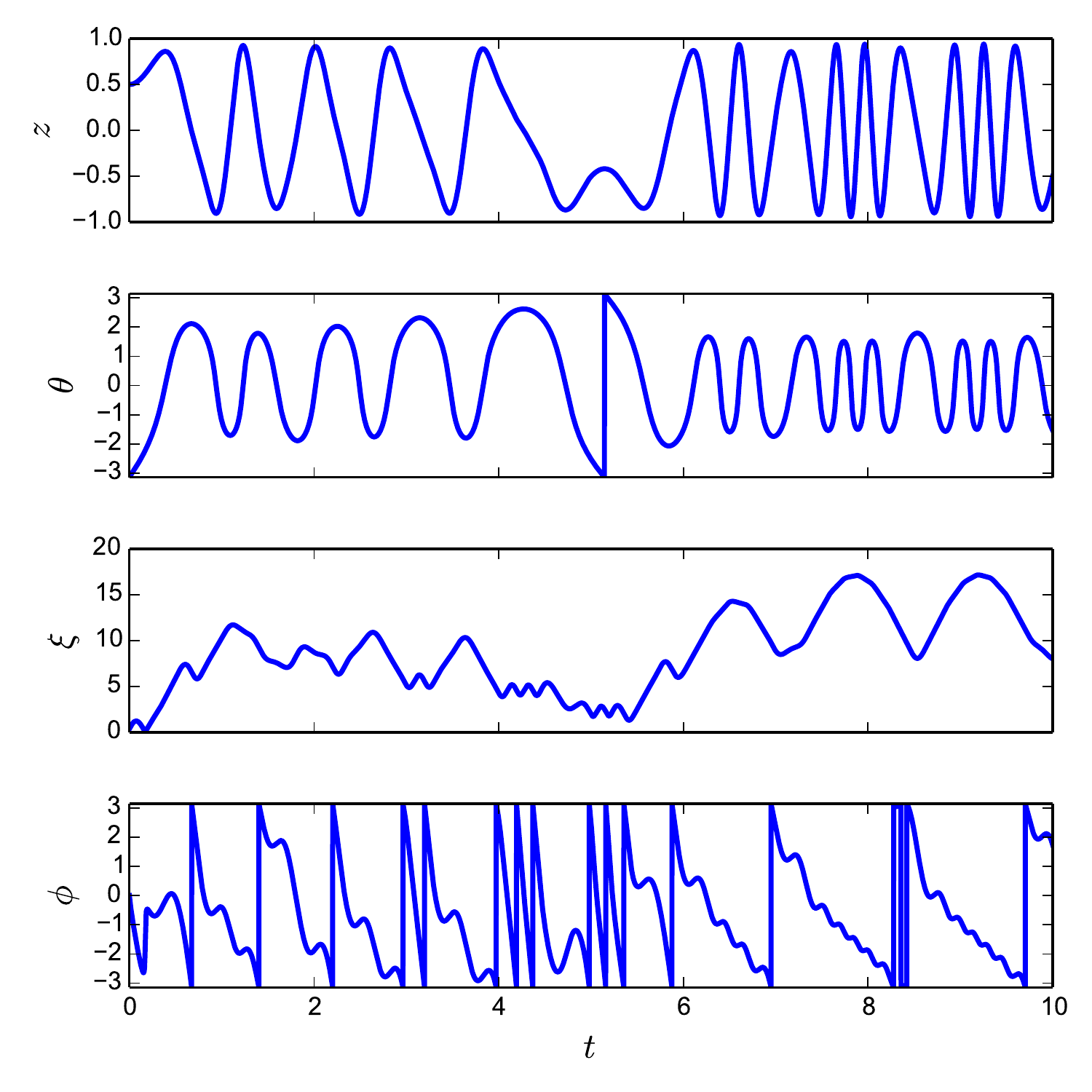}
\caption{(Color online) Time evolution of the variables $(z,\theta,\xi,\phi)$ started from the initial condition ${\bf X}(t=0)=(0.5,-\pi,0,0)$, which would correspond to a self-trapped, running phase solution in the case of a BJJ. Time is measured in units of $\hbar/J$. During the time evolution the population imbalance takes almost all possible values between $-1$ and $1$.  This solution is in sharp contrast to the self-trapping oscillations of pure BJJs.  The parameters are the same as for Fig. \ref{fig:Josc}.}
\label{fig:antiselftr}
\end{figure}

\begin{figure*}
\centering
\includegraphics[scale=0.6]{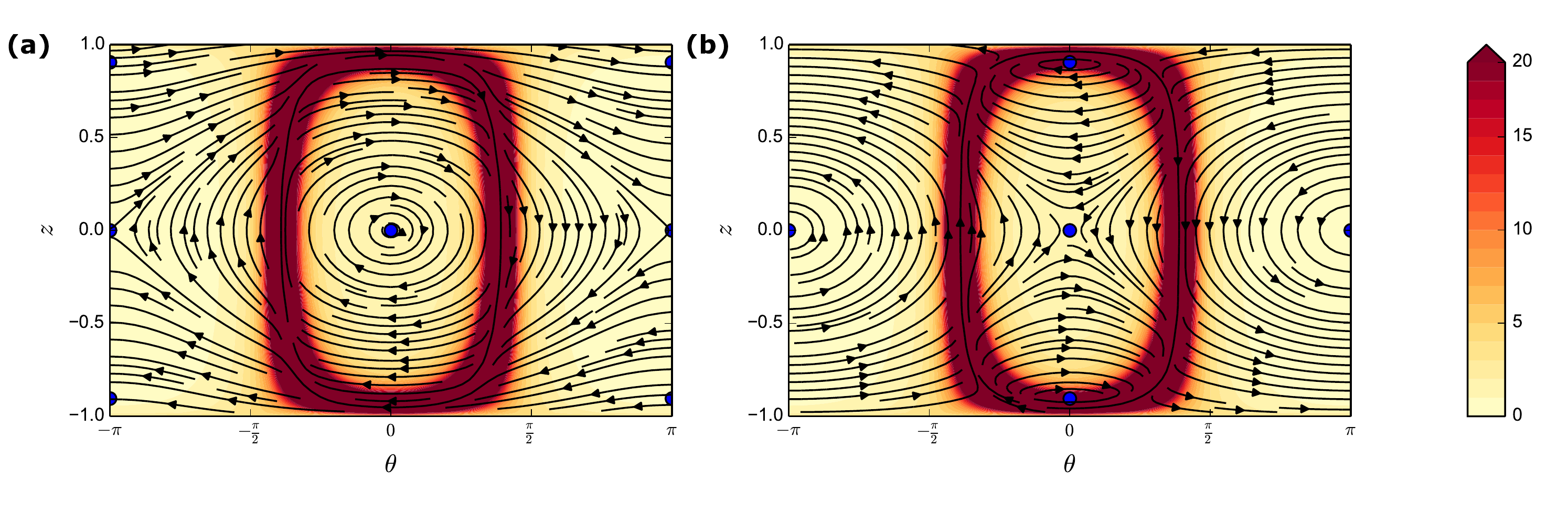}
\caption{(Color online) The phase space portrait, fixed points and trajectories of the BJJ model after adiabatically eliminating the cavity field. The coloring corresponds to the photon number $\bar\xi^2$. There is a ring shaped curve, where the photon number diverges. Outside the ring the effective cavity detuning is negative, while inside the detuning is positive. The parameters are the same as for Fig. \ref{fig:Josc} except that in panel b) the interaction is attractive $U<0$.}
\label{fig:phase_space}
\end{figure*}

Now let us turn to the \textit{finite imbalance equilibrium} solutions, featuring $0<|z|\le1$. Among the stationary points we study first those with $\theta=0$. In the bare atomic junctions, the ODEs admit these stationary points only when the interatomic interaction is attractive, $U<0$. For the light-matter interaction-dressed BJJs the situation is different. In fact, $\nu=J-W_{12}\xi^2$ can, in principle, change its sign for large enough photon number, as was shown for the simple laser-created hole potential in \cite{szirmai2014effect}.  
From Eq. (\ref{eq:odetheta}) we see that the fixed points with a finite imbalance and zero relative phase exist both for repulsive atom-atom interactions $U>0$ (when $J<W_{12}\xi^2$) and attractive atom-atom  interactions $U<0$ (when $J>W_{12} \xi^2$) interatomic interactions. 
We get the stationary points:
\begin{subequations}
\begin{align}
\label{fpnzz1}
{\bf X}_5
&=\left(\bar z,0,\frac{\hbar \eta}{\hbar\,\Delta_C-N_A\,
\big(W_0+W_{12}\sqrt{1-\bar z^2}\big)},\frac{\pi}{2}\right),\\
{\bf X}_6
&=\left(\bar z,0,\frac{\hbar \eta}{N_A\,\big(W_0+W_{12}\sqrt{1-\bar z^2}\big)
-\hbar\,\Delta_C},-\frac{\pi}{2}\right),
\end{align}
and similarly for $\theta=\pi$, one gets
\begin{align}
\label{fpnzz2}
{\bf X}_7&=\left(\bar z,\pi,\frac{\hbar \eta}{\hbar\,\Delta_C-N_A\,
\big(W_0-W_{12}\sqrt{1-\bar z^2}\big)},\frac{\pi}{2}\right),\\
{\bf X}_8&=\left(\bar z,\pi,\frac{\hbar \eta}{N_A\,\big(W_0-W_{12}\sqrt{1-\bar z^2}\big)
-\hbar\,\Delta_C},-\frac{\pi}{2}\right).
\end{align}
\end{subequations}
Here 
\begin{equation}
\label{barz}
\bar z=\pm \sqrt{1-\bigg(\frac{2(J-W_{12}\,\bar \xi^2)}{UN_A}\bigg)^2}.
\end{equation}
In Eq. \eqref{barz} $\bar \xi$ is a formal shorthand notation referring to the third component of the vectors ${\bf X}_5\ldots{\bf X}_8$.
Thus, Eq. \eqref{barz} is a non-linear equation for $\bar z$. More precisely, it is a cubic equation for $\sqrt{1-z^2}$ and therefore has either 1 or 3 real solutions. The conditions for $\bar z$ to be real are: $-UN_A/2\le (J-W_{12}\xi^2)\le UN_A/2$ for repulsive interactions ($U>0$), and $(J-W_{12}\xi^2)\ge -UN_A/2$ for attractive interactions ($U<0$). As $\xi$ has to be positive, it follows that for $\delta_C>0$ the two finite imbalance fixed points are ${\bf X}_5$ and ${\bf X}_7$ , while for $\delta_C<0$ the finite imbalance fixed points are ${\bf X}_6$ and ${\bf X}_8$. Oscillations around these fixed points are plotted in Fig. \ref{fig:antiselftr}. 

\section{Adiabatic elimination of photon dynamics}

The characteristic frequency of photon dynamics is the effective detuning, $\delta_C$, which is usually orders of magnitude larger than $\hbar^{-1}\nu$ characteristic for tunneling processes, $\hbar\delta_C\gg \nu$ \cite{ritch2013cold}. In this limit the time scales of photon and atom dynamics separate and the photon field can be adiabatically eliminated. The mean photon number is expressed as $\bar \xi=\eta/|\delta_C|$. When this is substituted into the ODEs \eqref{odes} one gets a modified set of BJJ equations:
\begin{subequations}
\label{eqs:odes2}
\begin{align}
\dot{z} &=-2\bar\nu\sqrt{1-z^2}\sin\theta,\label{eq:odez2}\\
\dot{\theta} &= \left(\tilde g + \frac{2\,\bar\nu}{\sqrt{1-z^2}}\cos{\theta}\right)z,\label{eq:odetheta2}
\end{align}
\end{subequations}
with $\bar\nu=(J-W_{12}\bar\xi^2)=(J-W_{12}\eta^2/\delta_C^2)$. The fixed points of Eqs. \eqref{eqs:odes2} are identical to the first two components of the fixed points of Eqs. \eqref{odes}, since the adiabatically eliminated photon number and phase, $\bar \xi$ and $\bar\phi$ at the fixed points of Eq. \eqref{eqs:odes2} are the same as the last two components of the fixed points of \eqref{odes}. Therefore, we have the zero imbalance fixed points: $(z=0,\theta=0)$ and $(z=0,\theta=\pi)$, and also the finite imbalance fixed points:  $(z=\bar z, \theta=0)$, with $\bar z$ being the solution of  Eq. \eqref{barz}, for $\tilde g\bar\nu>0$, and $(z=\bar z, \theta=\pi)$ for $\tilde g\bar\nu<0$.  Equation \eqref{barz} is a cubic equation here too with the same number of solutions as before.  The phase space portrait  of Eqs. \eqref{eqs:odes2} is illustrated in Fig. \ref{fig:phase_space} for 2 specific parameter settings. The trajectories show some similarity to the ones encountered in the pure BJJ problem \cite{raghavan1999coherent,albiez2005direct}. The shading of the plot is according to the cavity photon number, which is proportional to the inverse of the cavity detuning.

In Fig. \ref{fig:phase_space}, there is an oval shape around the origin, where $\delta_C=0$. Here both the photon number $\mathop{\bar\xi}^2$ and the effective tunneling $\bar\nu$ diverge.  This curve is a trajectory of the dynamics. It can be seen by first calculating the gradient vector of $\delta_C$ with respect to the variables $z$ and $\theta$ and then noticing that it is just orthogonal to the tangent of the trajectory [rhs of Eqs. \eqref{eqs:odes2}] when $\bar\nu$ diverges. As $\delta_C=0$ is a closed trajectory, it cuts the phase space into two parts. Every trajectory starting from its interior remains inside. It is very interesting to study the trajectory corresponding to initial conditions which are inside the $\delta_C=0$ curve but correspond to running-phase solutions of the pure BJJ problem (see Fig. 5 of Ref. \cite{raghavan1999coherent}). With cavity assisted tunneling these curves remain trapped inside the oval shaped region and become Josephson oscillations themselves. Consequently, every trajectory which starts from an initial condition of the exterior of the oval shape remains outside during time evolution. Therefore, the $\delta_C=0$ curve is a new separatrix in phase space. The separatrix of the pure BJJ dynamics is shattered when it crosses the $\delta_C=0$ curve. This explains why the self-trapping solutions are absent in Fig. \ref{fig:antiselftr}. To be more specific, the trajectory in Fig. \ref{fig:phase_space} (a) corresponding to Fig. \ref{fig:antiselftr} starts at the left border at $(z=0.5,\theta=-\pi)$. During time evolution, it gets closer and closer to the $\delta_C=0$ curve; therefore the photon number increases together with the effective tunneling rate. Close to the separatrix interaction is negligible compared to the effective tunneling therefore the trajectory follows to be close to the separatrix and eventually crosses the $z=0$ line. 

In the vicinity of this oval curve $\delta_C$ becomes small, the characteristic frequency of the photon dynamics slows down, and of course, we leave the regime where adiabatic elimination of the photon field can be performed. Nevertheless, apart from the vicinity of this region adiabatic elimination works, and the trajectories of Eq. \eqref{odes} more or less follow the phase space trajectories of \eqref{eqs:odes2}, which are  depicted in Fig. \ref{fig:phase_space}.

\section{Summary}

In summary, we have studied the coupled dynamics of a bosonic Josephson junction and the single mode of a high-Q optical cavity, when the cavity was arranged in such a way that cavity assisted tunneling could be relevant. The addition of the cavity mode to the bosonic Josephson junction is similar to the addition of a second mass to the rigid pendulum thus introducing another characteristic time.  We have shown that when the cavity can go close to resonance the dynamics changes and the self-trapped running phase solutions give way to more complex trajectories. The emergence of chaos in this system and the consequence of photon loss can be the subject of further studies.

The analysis presented so far paves the way to the quantum study of BJJs in the presence of an optical cavity mode. The dressed two-mode Bose-Hubbard Hamiltonian will feature different ground-states depending on the ratio of the on-site interaction to the assisted hopping. In such a context, an interesting issue is  the possibility for the Schr\"odinger-cat-like state of setting up as ground state by making negative the assisted hopping and keeping positive the on-site interaction  by-passing thus the collapse of the bosonic atomic cloud.

\section{Acknowledgements}

We acknowledge useful discussion with Jonas Larson. GSZ acknowledges support from the Hungarian National Office for Research and Technology under the contract ERC\_HU\_09 OPTOMECH, the Hungarian Academy of Sciences (Lend\"ulet Program, LP2011-016), the Hungarian Scientific Research Fund (grant no. PD104652) and the J\'anos Bolyai Scholarship. GM and LS acknowledge financial support from Universit\`a di Padova (Progetto di Ateneo grant No. CPDA 118083), Cariparo Foundation (Eccellenza grant 2011/2012), and MIUR (PRIN grant No. 2010LLKJBX).




\begin{thebibliography}{99}



\bibitem{raghavan1999coherent} S. Raghavan, A. Smerzi, S. Fantoni, and S.R. Shenoy, 
Phys. Rev. A {\bf 59}, 620 (1999). 

%
\bibitem{andrews1997observation} M. R. Andrews, C. G. Townsend, H.-J. Miesner, D. S. Durfee, D. M. Kurn, and W. Ketterle, Science {\bf 275}, 637 (1997).

\bibitem{wang2005atom} Y.-J. Wang, D. Z. Anderson, V. M. Bright, E. A. Cornell, Q. Diot, T. Kishimoto, M. Prentiss, R. A. Saravanan, S. R. Segal, and S. Wu, Phys. Rev. Lett. {\bf 94}, 090405 (2005).

\bibitem{albiez2005direct} M. Albiez, R. Gati, J. Folling, S. Hunsmann, M. Cristiani, and M. K. Oberthaler, Phys. Rev. Lett. {\bf 95}, 010402 (2005).

\bibitem{esteve2008squeezing} J. Estève, C. Gross, A. Weller, S. Giovanazzi, and M. K. Oberthaler, Nature (London) {\bf 455}, 1216 (2008).

\bibitem{leblanc2011dynamics} L. J. LeBlanc, A. B. Bardon, J. McKeever, M. H. T. Extavour, D. Jervis, J. H. Thywissen, F. Piazza, and A. Smerzi, Phys. Rev. Lett {\bf 106}, 025302 (2011).
%

\bibitem{mb} M. Lewenstein, Li You, J. Cooper, and K. Burnett,
Phys. Rev. A {\bf 50}, 2207 (1994). 

\bibitem{gio1} G. Mazzarella, M. Moratti, L. Salasnich, M. Salerno, 
and F. Toigo, J. Phys. B: At. Mol. Opt. Phys. 
{\bf 42}, 125301 (2009). 

\bibitem{gio2} G. Mazzarella, M. Moratti, L. Salasnich, and F. Toigo,
J. Phys. B: At. Mol. Opt. Phys. {\bf 43}, 065303 (2010). 

\bibitem{gio3} G. Mazzarella and L. Salasnich, Phys. Rev. A {\bf 82}, 033611 
(2010). 

\bibitem{gio4} G. Mazzarella, B. Malomed, L. Salasnich, M. Salerno, 
and F. Toigo, J. Phys. B: At. Mol. Opt. Phys. {\bf 44}, 035301 (2011). 

\bibitem{gio5} G. Mazzarella, L. Salasnich, A. Parola, and F. Toigo, 
Phys. Rev. A {\bf 83}, 053607 (2011). 

\bibitem{gio6} G. Mazzarella, L. Salasnich, and F. Toigo, 
J. Phys. B: At. Mol. Opt. Phys. {\bf 45}, 185301 (2010). 

\bibitem{julia2010bose} B. Juli\'a-D\'iaz, J. Martorell, and A. Polls,
Phys. Rev. A {\bf 81}, 063625 (2010).

\bibitem{julia2010macroscopic} B. Juli\'a-D\'iaz, D. Dagnino, M. Lewenstein, J. Martorell, and A. Polls
Phys. Rev. A {\bf 81}, 023615 (2010).


\bibitem{szirmai2014effect} G. Szirmai, G. Mazzarella, and L. Salasnich,
Phys. Rev. A {\bf 90}, 013607 (2014). 

\bibitem{ritch2013cold} H. Ritsch, P. Domokos, F. Brennecke, and T. Esslinger, Rev. Mod. Phys. {\bf 85}, 553 (2013).

\bibitem{maschler2005cold} C. Maschler and H. Ritsch, Phys. Rev. Lett. {\bf 95}, 260401 (2005).
\bibitem{larson2008mott} J. Larson, B. Damski, G. Morigi, and M. Lewenstein, Phys. Rev. Lett. {\bf 100}, 050401 (2008).
\bibitem{larson2008quantum} J. Larson, S. Fernandez-Vidal, G. Morigi, and M. Lewenstein, New J. Phys. {\bf 10}, 045002 (2008).
\bibitem{maschler2008ultracold} C. Maschler, I. B. Mekhov, and H. Ritsch, Eur. Phys. J. D {\bf 46}, 545 (2008).
\bibitem{vukics2009cavity} A. Vukics, W. Niedenzu, and H. Ritsch, Phys. Rev. A {\bf 79}, 013828 (2009).
\bibitem{niedenzu2010microscopic} W. Niedenzu, R. Schulze, A. Vukics, and H. Ritsch, Phys. Rev. A {\bf 82}, 043605 (2010).
\bibitem{fernandez2010quantum} S. Fernandez-Vidal, G. De Chiara, J. Larson, and G. Morigi, Phys. Rev. A {\bf 81}, 043407 (2010).
\bibitem{li2013lattice} Y. Li, L. He, and W. Hofstetter, Phys. Rev. A {\bf 87}, 051604 (2013).
%

\bibitem{corney1998homodyne} J. F. Corney, and G. J. Milburn, Phys. Rev. A {\bf 58}, 2399 (1998).
\bibitem{vukics2007microscopic} A. Vukics, C. Maschler, and H. Ritsch, New J. Phys. {\bf 9}, 255 (2007).
\bibitem{zhang2008cavity} J. M. Zhang, W. M. Liu, and D. L. Zhou, Phys. Rev. A {\bf 77}, 033620 (2008).
\bibitem{zhang2008mean} J. M. Zhang, W. M. Liu, and D. L. Zhou, Phys. Rev. A {\bf 78}, 043618 (2008).
\bibitem{zuppardo2014cavity} M. Zuppardo, J. P. Santos, G. De Chiara, M. Paternostro, F. L. Semiao, and G. M. Palma, preprint, arXiv:1407.8077  (2014).
\bibitem{larson2010ultracold} J. Larson, and J.P. Martikainen, Phys. Rev. A {\bf 82}, 033606 (2010).
%

\bibitem{rinck2011effects} M. Rinck and C. Bruder, Phys. Rev. A {\bf 83}, 023608 (2011)
\bibitem{mulansky2011impurity} F. Mulansky, J. Mumford, and D. H. J. O'Dell, Phys. Rev. A {\bf 84}, 063602 (2011).
\bibitem{mumford2014impurity} J. Mumford, J. Larson, and D. H. J. O'Dell, Phys. Rev. A {\bf 89}, 023620 (2014).
%

\bibitem{emary2003chaos} C. Emary and T. Brandes, Phys. Rev. E 67, 066203 (2003).
\bibitem{baumann2010dicke} K. Baumann, C. Guerlin, F. Brennecke, and T. Esslinger, Nature (London) {\bf 464}, 1301 (2010).
\bibitem{nagy2010dicke} D. Nagy, G. K{\'o}nya, G. Szirmai, G and P. Domokos,  Phys. Rev. Lett. {\bf 104}, 130401 (2010). 
\bibitem{zhou2014josephson} P. Zou, and F. Dalfovo, preprint, arXiv:1401.2007 (2014).
\bibitem{keeling2014fermionic} J. Keeling, M. J. Bhaseen, and B. D. Simons, Phys. Rev. Lett. {\bf 112}, 143002 (2014).
\bibitem{piazza2014umklapp} F. Piazza, and P. Strack,  Phys. Rev. Lett. {\bf 112}, 143003 (2014).
\bibitem{chen2014superradiance} Y. Chen, Z. Yu, and H. Zhai , Phys. Rev. Lett. {\bf 112}, 143004 (2014).
%

%
\bibitem{gerbier2010gauge} F. Gerbier and J. Dalibard,  New J. Phys. {\bf 12}, 033007 (2010). 
\bibitem{standardquantumopticsbook} W.P. Schleich, Quantum Optics in Phase Space (Wiley, Berlin, 2001).  

\end{thebibliography}
\end{document}